# Simplifying the vacuum bazooka

Junghwan Lee[1], Woong Sung Lee[2], Eunsoo Shin[3]

[1,2,3]*CheongShim International Academy, Gyeonggi-do, Republic of Korea*

[1]leejohn24@csia.hs.kr, [2]woongsunglee@csia.hs.kr, [3]eshin2020@csia.hs.kr

**Abstract**

This paper provides a simplified explanation of the vacuum bazooka through diagrams and builds a theoretical model only using concepts found in introductory mechanics. Our theory suggests that the velocity of the projectile is proportional to the hyperbolic tangent of time, and experimental measurements support this claim. We also find that the vacuum bazooka could be used to demonstrate the concept of terminal velocity in a classroom setting.

## 1. Introduction

The vacuum bazooka, also known as vacuum cannon, can be built with simple plastic pipe, a light projectile, and a vacuum cleaner. Though a simple device, the vacuum bazooka has proved to be an exciting demonstration of mechanics, and have drawn the attention of physicists and engineers across the world [1-5].

Cockman made an early attempt to propose a more convenient version of the vacuum bazooka [1], and several technical improvements later followed the research. For example, a common misconception that the projectile reaches the speed of sound was clarified when Ayars has proved that there is a maximum kinetic energy that the projectile can reach [2].

There were improvements to research methods as well. Peterson has brought in Schlieren Photography to describe the muzzle velocity of the projectile [3]. Additionally, French has used CFD (Computational Fluid Dynamics) analysis to explain the air flow within the vacuum bazooka [4].

Though previous papers have detailed the phenomenon well, the vacuum bazooka is meant to be an entertaining physics experiment for younger audiences, and we believed that there could be a more fundamental approach to understanding the device. Therefore, this paper seeks to simplify the vacuum bazooka.

## 2. Phenomenon and Theory

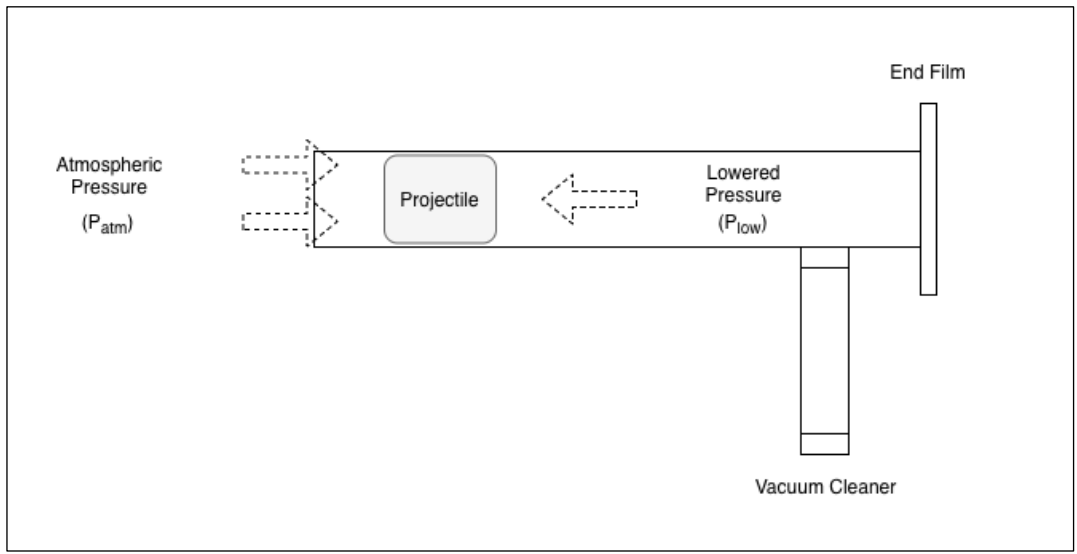

**Figure 1:** A schematic diagram of the vacuum bazooka. The arrows represent the difference in pressure on the projectile.

A vacuum bazooka set-up consists of a pipe, projectile, vacuum cleaner, and an end film (Figure 1). A hole close to the end of the tube is first made, and the vacuum cleaner is attached to that hole. The end of the tube that adjacent to the vacuum cleaner hole is then covered with an end film, which can be aluminum foil or paper. A projectile with a cross-sectional area similar to the cross-sectional area of the tube is then inserted into the tube. The projectile then speeds through the pipe, breaks the end film, and bursts out.

This is because when the projectile and the end film cover the ends of the pipe, the vacuum cleaner lowers the pressure inside the tube. Then, the pressure acting on the front of the projectile (lowered because of the vacuum cleaner) becomes less than the pressure acting on the back of the projectile (atmospheric pressure), and this difference propels the projectile forward.

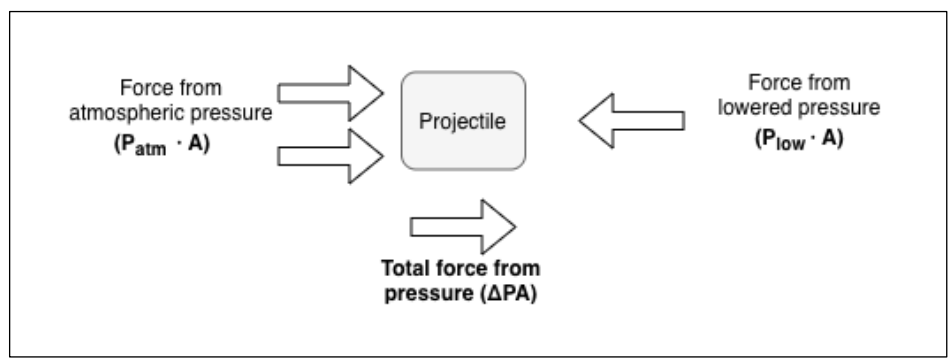

**Figure 2:** The forces that act on the projectile due to the pressure difference.

The force caused by pressure difference ($F_P$), can be modeled as

$$F_P = \Delta P A \tag{1}$$

where $\Delta P$ is the difference between the atmospheric pressure and the air pressure in pipe and $A$ is the cross-sectional area of the projectile.

When we observed the projectile's motion inside the tube, we realized that a thin layer of air is created between the pipe and the projectile as stated in previous researches [4,5]. This layer causes the projectile to float as it progresses through the tube, so there is no frictional force between the projectile and the tube.

However, there still exists a force that opposes the projectile's acceleration, and that is the drag force caused by air resistance. Though an ideal vacuum bazooka would make a complete vacuum inside the tube, vacuum cleaners can only lower the pressure inside the tube to about 80% of its original value, and therefore, air is present inside the tube.

The force caused by air resistance ($F_R$) can be modeled as

$$F_R = kv^2 \tag{2}$$

where $k$ is the coefficient of resistance and $v$ is the velocity of the projectile.

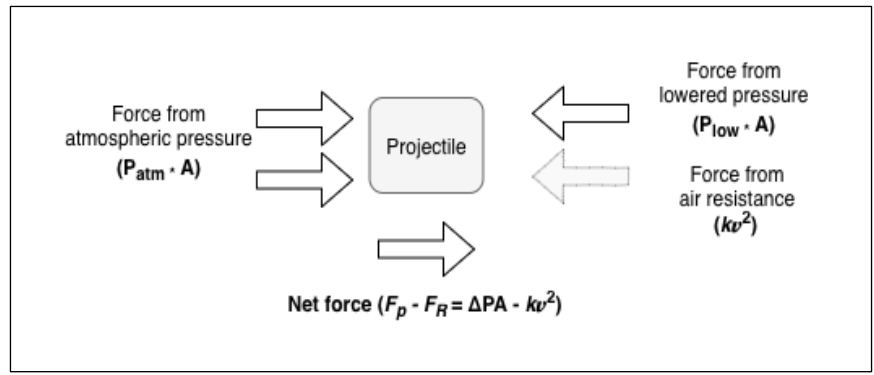

**Figure 3:** The forces that act on the projectile. Modeled in Equation (3).

If $F$ is the net force acting on the projectile, the relationship between the two forces in Equation (1) and (2) can then be expressed into

$$F = F_P - F_R \tag{3}$$

Substituting equation (3) with equations (1) and (2) then gives us

$$m\frac{dv}{dt} = \Delta P A - kv^2 \tag{4}$$

Then, solving Equation (4) for the velocity of the projectile (v) gives us our final equation

$$v = \sqrt{\frac{\Delta PA}{k}} \tanh(\frac{\sqrt{k\Delta PA}}{m}t) \qquad (5)$$

Though the value of $k$ is unknown, Equation (5) suggests that the velocity of the projectile is proportional to the hyperbolic tangent of time.

## 3. Experiment

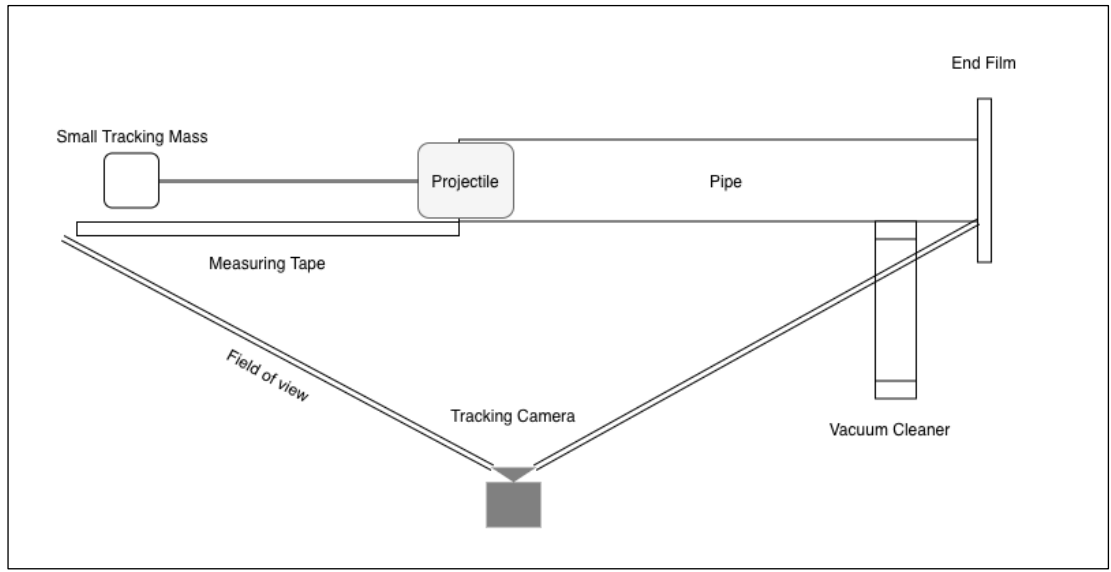

**Figure 4:** A schematic diagram of the experimental setup

To test if the relationship suggested in Equation (5) is correct, we set-up an experiment as shown in Figure 4. The apparatus includes a small tracking object (Styrofoam ball of radius 0.5 cm, colored red) that is attached to the projectile with a thin string. As the projectile is released, the tracking object was tracked using a camera at 50 frames per second and analyzed through the tracker software. The measuring tape was used to scale the lengths taken in the video.

Since the mass of the tracking object was 0.16 grams, we considered it negligible compared to the mass of the projectile (60 grams). In addition, the Styrofoam ball floated in the air as the projectile accelerated through the pipe, so there was no frictional force between the tracking object and the underlying surface.

Our vacuum bazooka was constructed with a PVC pipe as the tube and a plastic cylinder as the projectile. The pipe was 3 meters long with a radius of 2 cm, and the plastic cylinder was 12 cm long with a radius of 1.98 cm. The mass of the plastic cylinder was 60 grams. The vacuum cleaner that we used was able to create a pressure difference of 24000 Pascals, as measured with a PASCO Dual Pressure Sensor.

We shot the projectile 20 times and recorded the velocity of the projectile using the tracker software. The average of the 20 trials was drawn into a graph with its standard error as the error bar, and a hyperbolic tangent fit was found using the scipy.optimize.curve_fit function in Python(scipy). The tanh(x) fit function was then drawn onto the data points using OriginPro 8 to see if the projectile's velocity is proportional to the hyperbolic tangent of time (Figure 5).

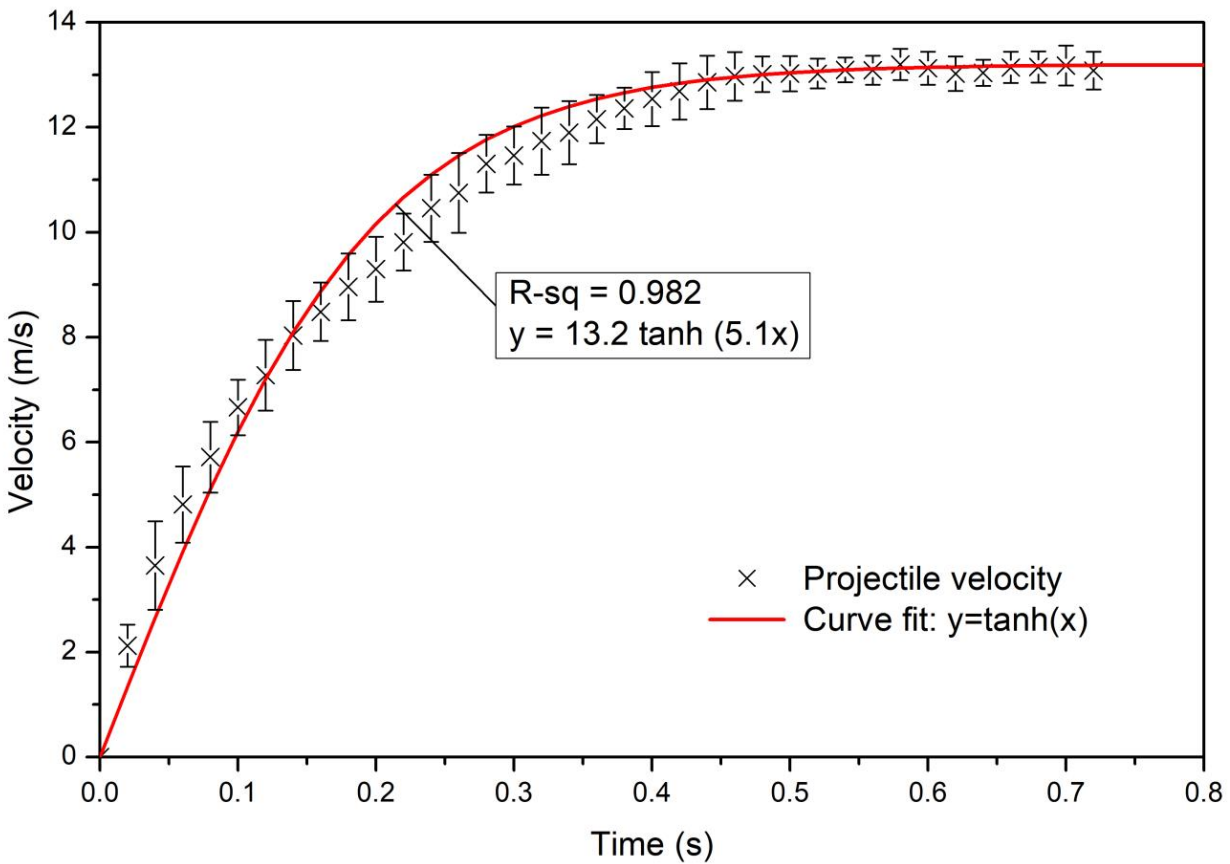

**Figure 5:** The velocity of the projectile versus time, with a fitted hyperbolic tangent function

There are some variances in the trial because we used our hands to release the projectile inside the tube. Lifting our hands off the projectile might have given a slight force that made the projectile unstable at first. However, as time passes and the projectile moves through the pipe, the projectile seems to recover from the effect of our hand and exhibit a more stable motion, leading to less variance (smaller error bar).

Nevertheless, the fitted function showed almost no deviation from the error bounds, and the R-squared value of the fitted function was 0.982, which indicate that the relationship between projectile velocity and time was well represented through Equation (5). We have attempted to fit other functions with similar tendency as our data points, such as sqrt(x) and log(x) function; however, their R-squared values were lower than that of the tanh(x) function.

## 4. Conclusion

As an attempt to provide a more straightforward approach to the vacuum bazooka, we have designed a theoretical model that predicts the tendencies of the projectile's motion using only fundamental concepts. Though our final equation involved a coefficient of resistance that remains unknown, our experiments validated that our model is adequate to explain the tendency of the projectile's motion in the vacuum bazooka. Those who wish to grasp the idea of a vacuum bazooka and how it works could use this paper to understand the basic principles involved in this phenomenon.

It is worth noting that our model for the projectile's velocity is analogous to the velocity of a free-falling object. Just like how air resistance opposes the pressure force in the vacuum bazooka, the air resistance opposes the gravitational force on a free-falling object as well.

The only difference is that the net force on the vacuum bazooka projectile is determined horizontally by $\Delta PA$ while the net force on the free-falling object is determined vertically by $mg$. Therefore, in both phenomena, the object in motion reaches a terminal velocity due to drag force.

With that note, the vacuum bazooka could perhaps be used as an apparatus to demonstrate the concept of terminal velocity and air resistance inside a classroom setting.